# Control Surfaces: Using the Commodore 64 and Analog Synthesizer to Expand Musical Boundaries

**Daniel McKemie**
Independent Artist
daniel.mckemie@gmail.com

## ABSTRACT

*Analog-digital hybrid electronic music systems once existed out of necessity in order to facilitate a flexible work environment for the creation of live computer music. As computational power increased with the development of faster microprocessors, the need for digital functionality with analog sound production decreased, with the computer becoming more capable of handling both tasks. Given the exclusivity of these systems and the relatively short time they were in use, the possibilities of such systems were hardly explored. The work of José Vicente Asuar best demonstrated a push for accessibility of such systems, but he never received the support of any institution in order to bring his machine widespread attention. Modeled after his approach, using a Commodore 64 (or freely available OS emulator) and analog modular hardware, this paper aims to fashion a system that is accessible, affordable, easy to use, educational, and musically rich in nature.*

## INTRODUCTION

The complicated discourse surrounding the often-divergent camps touting Analog over Digital, or vice versa, has found its way into many aspects of contemporary life, and in the field of music creation, this could not be more palpable, if not played out. Still, while there are certain computational elements that generally excel in analog circuitry over digital, with randomness being an arguably prominent one, algorithmic, compositional, and procedural processes can be declared and controlled in a computer much more efficiently than analog circuitry alone. By blending analog and digital systems, two seemingly disparate computational models, new possibilities emerge for the exploitation of modern electronic music techniques that not only retain the primary functionalities of each system but creates one that is wholly new.

## HISTORY

The PIPER 1 and GROOVE systems developed at the University of Toronto and Bell Labs, respectively, in the 1960s demonstrated the earliest realizations of analog-digital hybrid systems [1]. The limited power of computers at the time were unable to construct audio in real time and the solution provided by the PIPER and GROOVE systems was to have the computer *shape* the events while delegating sound production duties to the modular synthesizer. The GROOVE system expanded on the functionality of the PIPER by enabling the system to read and store performer's interactions with the potentiometers on the synthesizer. This introduced the possibility for human-interactions and performance to be stored, and conversely, for events to be programmed, recalled, and tooled in real-time [2].

In the following decade, and with no institutional support, Chilean composer José Vicente Asuar designed and built his COMDASUAR (COmputador Musical Digital Analogico Asuar) system, which consisted of an Intel 8080 microprocessor-based computer housed alongside a custom-built hardware synthesizer [3]. Unlike his North American predecessors' designs in the 1960s, in which the computer and analog synthesizer were independent machines, Asuar designed his to be one cohesive unit.

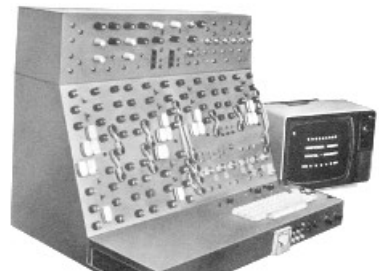

**Figure 1**. COMDASUAR system designed by Chilean composer José Vicente Asuar in 1978 for $1000 (approx. $4000 in 2020), a fraction of the price of similar systems at the time.

Twenty-six subprograms capable of constructing and morphing musical ideas were coded into the system using machine language. These subprograms enabled the user to declare/define a variety of musical parameters, such as pitch, duration, textures (in the form of glissando and vibrato), as well as transmutation of these musical parameters. Such transmutations included retrograding and transposing, to constructing canons of passages. Asuar also created what he called "heuristical" programs that allowed for non-performer interjection of the computer, such as probability distributions assignable to musical parameters [3, 4]. Another striking feature of Asuar's system was the ease by which musical events could be programmed. He approached it as a pedagogue and might have sought to include an effective programming tool that could be accessible to all levels of musicians [5].

He did not want the technological hurdle to create barriers to those wishing to learn and use his system. It was Asuar's approach to a hybrid system that was most influential in devising the one outlined in this paper. The Commodore 64 (C64) was released in 1982 and runs on the BASIC 2.0 operating system, a simple and elegant language designed for the non-engineer in mind [6, 7]. While the C64 fell out of favor in the 1990s in favor of faster machines running more user-friendly operating systems, it recently enjoyed a resurgence thanks in part to the rise of vintage gaming, reliable and multi-platform C64 emulators across several platforms, and "chiptune" music [8].

The system discussed herein, the synthesizer component will refer to any available brand or arrangement of modules that the user wishes to use, so long as the end user has a certain level of understanding of the fundamental elements of analog synthesis and BASIC programming, or the enthusiasm to explore further resources in which to learn.

## HARDWARE IMPLEMENTATION

### 1. Modular Synthesizer

There are two different types of signals in a modular synthesizer; audio and control voltage (CV). Audio signals include all audible frequencies, approximately 20Hz-20kHz. CV signals are inaudible, are generally below 20H, and are used to modulate audio signals or other CV signals. An example of an audio signal would be the tone that emerges from an oscillator, while a CV signal might be the output of an ADSR envelope, gates and triggers, or a series of stored voltages, ie. a sequencer [9].

Modular synthesizer signals are hotter, ie. higher voltage, than line level signals, a preamp is required to boost the signals coming from the C64 to a level that will be functional in the modular environment. Most envelope followers will provide this functionality locally, directly converting the line level of the C64 to proper levels for the synthesizer, but a variety of mixers, amplifiers, and adapters in can be used to achieve this signal level conversion.

### 2. C64 to the Modular Synthesizer

The C64 has two options for outputting audio and video signals: 5-pin DIN and radio frequency (RF). The single RF jack sends both audio and video in one cable and can be useful for establishing quick connections for computer work alone. Many televisions have RF inputs, but it was found that a much higher amount of noise and interference occurs between the audio and video paths and is not recommended for use when connecting to the synthesizer.

The 5-pin output splits the audio to dual mono and video. Custom cables are available through DIY outlets to convert the 5-pin to RCA, or through consulting schematics available online. In this case, a custom 5-pin to RCA cable is used as shown in Figure 2.

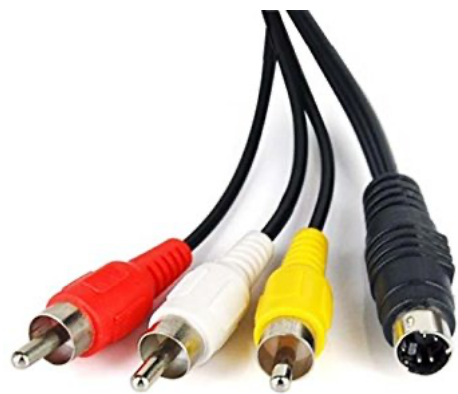

**Figure 2**. A custom made 5-pin DIN to RCA cable.

While the connection between the C64 and the synthesizer is effectively format and/or manufacturer-agnostic, it is essential that the synthesizer contain an envelope follower. Envelope followers (also known as amplitude followers or envelope detectors) are devices that measure the power of incoming audio signals and converts that signal to useable control voltages [10]. The audio output from the C64 will ultimately be translated into CV through the envelope follower, with necessary attention to some factors along the way.

Morton Subotnick's *Sidewinder* from 1971 employs the use of numerous envelope detectors and recorded pulses to act as control signal triggers. The approach illustrates that pulse wave material can be composed on the synthesizer (or by external means), and combined with multiple detectors to achieve resultant patterns to be routed anywhere in the system, including larger systems and lightning rigs [11]. The approach illustrated here is not so different, except the control source is that of a C64.

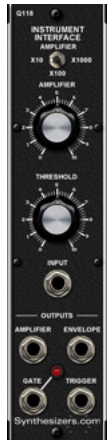

**Figure 3**. An out-of-the-box envelope follower, designed by Synthesizers.com.

There are four basic waveforms plus a noise source that can be output from the C64, and any of these sound generators will suffice. Every decision regarding waveform type, frequency, modulation type, and how the outgoing audio is filtered will contribute to how the envelope follower reacts to the incoming audio. While it should be expected that the behaviors and features of a hardware envelope follower will vary from module to module, including two modules of the same make and model, the basic functionality should be obtainable, even with the simplest of designs. The signal from the computer through the envelope follower(s) can now be routed, multiplied, modulated, and treated in any number of ways, limited only by the user's hardware and creativity.

While this paper is not intended to focus on the specific modules and brands, given the relative ubiquity of these technologies, it is relevant to state that few modules have the capability to execute the unique functionalities of a standalone computer, regardless of its vintage. There does seem to be a growing interest in the implementation of more *computer-based* modules with CPU cores mounted to a hardware interface, or modules built and marketed as programmable digital signal processors, but with few exceptions, the possibility for the code-based control of the modular synthesizer is still within the domain of the standalone computer.

## SOFTWARE IMPLEMENTATION

### 1. Programming in BASIC

The BASIC (Beginners' All-purpose Symbolic Instruction Code) family of languages has many of the same features of modern languages such as counters, if/then statements, and built-in arithmetic and mathematical operations. These points make it a practical entry level environment for those wishing to learn fundamental programming techniques. No third-party programs are necessary to utilize the C64 as work can be done directly in the BASIC 2.0 OS using only the language. There is extensive documentation available online to learn BASIC, and it is very much encouraged for both beginner and advanced programmers to pursue this path. It is easy and fun to learn, can serve as a great introduction to programming, and also reinforce programming fundamentals.

### 2. SID Chip

As is the case with programming in BASIC, no third-party programs are necessary on the C64 to achieve musical results as this can be done by programming the memory addresses of the SID Chip directly using the POKE function. Each address on the chip controls one of the parameters available on it.

As stated before, the SID chip is equipped with three voices each with four waveforms (sawtooth, triangle, pulse, and sine) with oscillator sync capabilities, noise generators, amplitude modulation, ring modulation, combination LP/HP/BP filter with variable resonance, and ADSR generators. These features are spread across the three voices with some minor limitations arising depending on how the voices are modulated together. The user can store the SID's base address of 54272 into a variable which provides easier access to all the parameters housed on the chip.

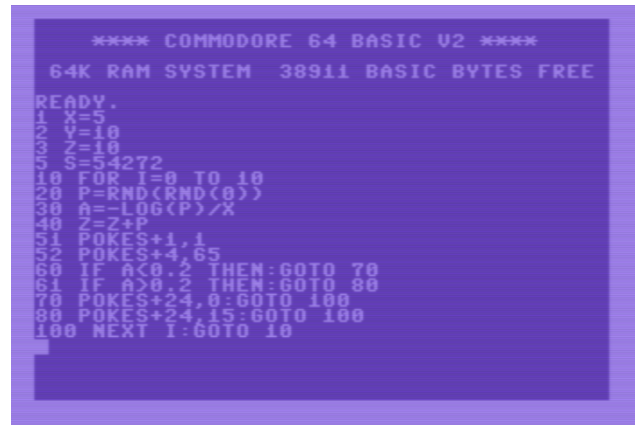

**Figure 4**. BASIC example of accessing SID addresses, assigning values, writing conditionals, and looping.

The frequency values of the oscillators are programmed directly on two addresses called the high and low bits. While seamlessly sweeping frequencies is not a strong area of the C64's aesthetic capability, microtonality and composing with alternative tuning systems is very much an option. The oscillators and noise source are available with options to synchronize and equipped with a ring modulator, all programmable independently on each voice respectively. The pulse wave allows for eight different degrees of pulse width to be changed on its own set of addresses.

Coupled with each oscillator is an envelope generator. This is considered one of the strongest points of the SID chip. These envelopes amplitude modulate each oscillator before being output to a primary filter. Each stage ranges from a couple of milliseconds up to 24 seconds as shown in Table 1, and the manner in which they are declared demonstrates a workflow unique to the SID [6].

| Value | Attack Rate | Decay/Release |
|-------|-------------|---------------|
| 0 | 2 ms | 6 ms |
| 1 | 8 ms | 24 ms |
| 2 | 16 ms | 48 ms |
| 3 | 24 ms | 72 ms |
| 4 | 38 ms | 114 ms |
| 5 | 56 ms | 168 ms |
| 6 | 68 ms | 204 ms |
| … | … | … |
| 13 | 3 s | 9 s |
| 14 | 5 s | 15 s |
| 15 | 8 s | 24 s |

**Table 1**. Bit value to envelope times of the SID chip

As seen in Table 1, stage lengths are fixed and called by declaring the respective bit on the chip. These same methods are employed when determining pitch, filter cutoff, etc., which allows algorithmic processes to move more seamlessly around the chip [5, 6]. While modern software enables more detailed customization of envelope stages and times, the flexibility of the SID in regard to digital audio in home computing was unprecedented at the time. Table 1 illustrates only a small range of the envelope generation abilities of the chip while the composer can navigate details in those timing structures and independently program each one.

The modulated sound sources are mixed together and routed through to a primary filter and volume nodes before being output as audio. High Pass, Low Pass, and Band Pass with control over cutoff and Q rate. The filter type and controls each have their own addresses. The filter on the SID chip provides a characteristic to the end result that is often shunned by analog purists and audiophiles, but at the same time provides so much of the character of the resulting sound. More on the musical uses of the filter and other SID parameters will be elaborated upon in the following section.

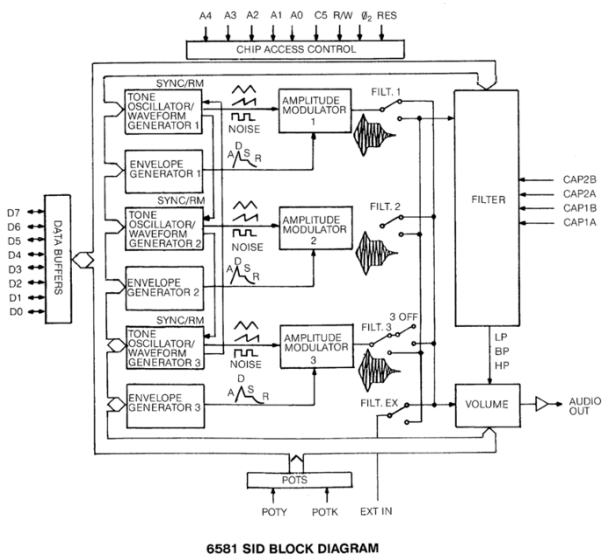

**Figure 5**. The SID chip feature and signal flow diagram.

# CHOICE MUSICAL APPROACHES

### 1. Gate and Trigger Control Signals

Trigger and gate signals are a foundational piece of controlling musical events on a modular synthesizer Triggers go from zero to a high voltage and back to zero quickly, while gates remain at a high voltage for a determined amount of time. These signals can be programmed to happen at choice times for decided lengths in BASIC and the composer may patch the synthesizer accordingly to produce musical events based on input of these signals. Taking the same approach as Asuar's system, programming pulse wave signals through an envelope follower yields quite satisfying results.

Routing this signal into an envelope follower gives us the desired signal level to be useable in a modular synthesizer environment. In our example, the gate signal is sent to a Sample and Hold module to trigger the hold of the incoming signal, which will be the amplified pulse wave from the C64. The output of the S&H will be patched into a low pass filter. Additionally, we will multiply the pulse wave signal and use it as an audio-rate modulator into one of the oscillators.

With a few lines of code and a just as many patch cables, a dual musical ecosystem was created. The composer may choose to further program more complex and algorithmic musical results on the C64 or devise a more involved synthesizer patch. Or most preferably, BOTH!

### 2. Audio Modulation

The most immediate and streamlined means of producing musical results within the synthesizer from the C64 is through audio modulation. While this does not always necessarily constitute 'controlling' the modular synthesizer by means of computer, it does provide an array of programmatically controlled colors routable to a hardware system. This technique is also available to anyone without the use of an envelope follower.

Audio modulation signals can be used the same as any other audio source whether it be native to the synthesizer or external. Some use cases include acting as a carrier frequency in mixing and ring modulation paths, a frequency modulation source into an oscillator, and as a standalone audio source to be treated similarly to other sound sources in the system.

### 3. Other Programs, Modern Developments, and MIDI

While the C64 has been out of production for almost three decades, the enthusiasm for them has not entirely disappeared. Hobbyists and professionals alike have developed emulators, new software programs, and even updated hardware devices to use with modern musical equipment. One example being the Mssiah Cartridge by 8bit Ventures. The cartridge comes equipped with a 5-pin MIDI input connection to control a handful of included programs [12]. The possibilities with such a device extend beyond the scope of this paper, but there is no shortage of MIDI/CV devices available on the market to further connect these two environments in engaging ways.

## DEPRECATION AND AN ARGUMENT IN FAVOR OF THE ARCHAIC

My decision to pursue the C64 over more contemporary computational tools is largely inspired by how its procedural simplicity can lead to so many interesting events and processes, and even though the model was discontinued in 1994, it is commonly more safeguarded against deprecation when compared to modern software. The features provided in the SID Chip are rich enough to provide extensive musical possibilities for the user while also serving as an introductory method to building and using hybrid systems. Entire compositions can be written and exchanged as text files and with the widespread availability of BASIC OS emulators, the original C64 is not even necessary. BASIC contains many of the key functionalities that modern programming languages have today, and this, translation is a relatively simple procedure. Pieces that utilize a hybrid system with a C64 emulator stand to outlast many modern software realizations written today, seeing as the BASIC OS stands frozen in time and widely supported. So long as quality emulators are available, and a diverse and engaged community continues to support this type of work, compositions of this sort will be easy to realize and preserve for future presentation. Additionally, despite the unfortunate reality of Mr. Asuar having to voluntarily shelve his project due to a lack of institutional support [5], it is encouraging to understand and explore his work as a use case of a seemingly deprecated technology that can still yield tremendous musical results.

## FUTURE WORK

While there is an abundance of information to be learned when revisiting vintage technology, the goal of this paper is not to make a case to live in the past and use dated techniques for music making. Rather, it has been my goal to illustrate that rich musical tools exist in places that may have not been appropriately recognized, like the C64 and this begs the question: What would the consumer electronic instrument look like had Mr. Asuar's creation been more widely supported and marketed? Would hybrid computer and synthesizer systems have become part of the norm? Would MIDI have happened, and if so, would the protocol have been different?

It is worth reiterating that because of emulators music made on the C64 may have a longer lifespan than some modern software. This is not to say that Max/MSP will not be supported in a decade's time, but updates and lack of support can cause patches to crash or cease to function as intended [13]. There are no new updates to the BASIC 2.0 OS but only to the emulators in which they run, and as long as emulators exist and are supported by a community including and beyond experimental music makers, the code will function as intended.

This point also stands as a sound argument for using the C64 in a pedagogical sense to teach both programming and computer music fundamentals, even if done so without the synthesizer or the original hardware computer itself. The BASIC language is simple and accessible across numerous platforms free of charge. This is the same set of standards and ideas that Mr. Asuar had for his system, which could have had a great impact on how the subject is taught had his machine at the time gotten the recognition it deserved.